\documentclass[aps,reprint,longbibliography] {revtex4-1}

\usepackage{amsmath,graphicx,amsfonts,amssymb}
\usepackage{times}
\usepackage[colorlinks=true,linkcolor=blue,urlcolor=blue,citecolor=blue,breaklinks=true]{hyperref}

\usepackage[utf8]{inputenc}
\DeclareUnicodeCharacter{03B4}{$\delta$}
\DeclareUnicodeCharacter{2009}{\,}
\DeclareUnicodeCharacter{03BC}{\mbox{$\mu$}}
\DeclareUnicodeCharacter{2215}{/}

\usepackage[english]{babel}

\makeatletter
\def\bbl@set@language#1{%
	\edef\languagename{%
		\ifnum\escapechar=\expandafter`\string#1\@empty
		\else\string#1\@empty\fi}%
	\@ifundefined{babel@language@alias@\languagename}{}{%
		\edef\languagename{\@nameuse{babel@language@alias@\languagename}}%
	}%
	\select@language{\languagename}%
	\expandafter\ifx\csname date\languagename\endcsname\relax\else
	\if@filesw
	\protected@write\@auxout{}{\string\select@language{\languagename}}%
	\bbl@for\bbl@tempa\BabelContentsFiles{%
		\addtocontents{\bbl@tempa}{\xstring\select@language{\languagename}}}%
	\bbl@usehooks{write}{}%
	\fi
	\fi}
\newcommand{\DeclareLanguageAlias}[2]{%
	\global\@namedef{babel@language@alias@#1}{#2}%
}
\makeatother

\DeclareLanguageAlias{en}{english}

\makeatletter
\def\@bibdataout@aps{%
	\immediate\write\@bibdataout{%
		@CONTROL{%
			apsrev41Control%
			\longbibliography@sw{%
				,author="08",editor="1",pages="1",title="0",year="1"%
			}{%
				,author="08",editor="1",pages="1",title="",year="1"%
			}%
		}%
	}%
	\if@filesw \immediate \write \@auxout {\string \citation {apsrev41Control}}\fi 
}
\makeatother

\def\eq#1{Eq.~(\ref{#1})}
\def\la{\langle}
\def\ra{\rangle}

\def\da{\dagger}

\date\today
\begin{document}
	
\title{Density profile of a semi-infinite one-dimensional Bose gas and bound states of the impurity} 

\author{Aleksandra Petkovi\'{c}}
\author{Benjamin Reichert}
\author{Zoran Ristivojevic}
\affiliation{Laboratoire de Physique Th\'{e}orique, Universit\'{e} de Toulouse, CNRS, UPS, 31062 Toulouse, France}
	
\begin{abstract}
We study the effect of the boundary on a system of weakly interacting bosons in one dimension. It strongly influences the boson density which is completely suppressed at the boundary position. Away from it, the density is depleted over the distances on the order of the healing length at the mean-field level. Quantum fluctuations modify the density profile considerably. The local density approaches the average one as an inverse square of the distance from the boundary. We calculate an analytic expression for the density profile at arbitrary separations from the boundary. We then consider the problem of localization of a foreign quantum particle (impurity) in the potential created by the inhomogeneous boson density. At the mean-field level, we find exact results for the energy spectrum of the bound states, the corresponding wave functions, and the condition for interaction-induced localization. The quantum contribution to the boson density gives rise to small corrections of the bound state energy levels. However, it is fundamentally important for the existence of a long-range Casimir-like interaction between the impurity and the boundary.
\end{abstract}

\maketitle

\section{Introduction}

Boundaries play an important role in one-dimensional quantum liquids affecting, e.g., correlation and response functions, as well as the ground-state energy \cite{PhysRevB.46.10866,eggert_impurities_1995,fabrizio_interacting_1995,PhysRevLett.76.1505,gaudin_boundary_1971,PhysRevLett.56.742,alcaraz_surface_1987, PhysRevLett.123.250602}. The boundary has also an impact on the density profile of particles $n(x)$, which is suppressed at the boundary position $x=0$. Away from it, the particle density in fermionic systems shows so-called Friedel oscillations \cite{friedel_metallic_1958}. They describe an oscillatory decay of $n(x)-n_0$, where $n_0$ is the average density. The envelop of Friedel oscillations follows $1/x^K$ law for spinless fermions  \cite{fabrizio_interacting_1995,PhysRevLett.75.3505}. Here $K$ is the Luttinger liquid parameter, which is determined by the interaction between fermions. The periodicity of oscillations is controlled by the average fermion density. 

Friedel oscillations in fermionic systems  are studied within the harmonic Tomonaga-Luttinger liquid description \cite{fabrizio_interacting_1995,PhysRevLett.75.3505}. This is the low-energy theory of both, interacting fermions and bosons in one dimension \cite{Haldane}.  It was applied for bosons in Ref.~\cite{Cazalilla_2002} where  the same pattern of density oscillations was found for separation longer than the inverse mean density of bosons, $1/n_0$. Applied to the special case of weakly-repulsive bosons where~\mbox{$K\gg 1$}, the result of Ref.~\cite{Cazalilla_2002} implies very rapid saturation of $n(x)-n_0$. Such fast recovery of the density is not expected to occur in weakly interacting superfluids where the density should change on the scale comparable to the healing length $\xi\sim K/n_0$, which is much longer than $1/n_0$. Thus, in order to describe the density profile of bosons one needs to go beyond the harmonic Tomonaga-Luttinger liquid theory.

In this paper we study weakly-interacting bosons in a semi-infinite system. We consider the equation of motion for the field operator. At the mean-field (classical) level it reduces to the Gross-Pitaevskii equation \cite{pitaevskii_bose-einstein_2003}. Its solution reveals that the spatial extent of the depletion of the boson density imposed by the boundary is controlled by the healing length $\xi$. At distances longer than $\xi$, the mean-field density exponentially rapidly reaches the constant value $n_0$.  Accounting for the effect of quantum fluctuations around the mean-field solution within the Bogoliubov-de Gennes formalism, we find that the density reaches $n_0$ much slower, following $1/x^2$ law. Our approach enables us to obtain an analytic expression for the density profile at all distances.

The nonuniform density profile of the Bose gas determines the potential for a weakly-coupled quantum impurity introduced in the system. For repulsive interaction between the impurity and the bosons, the impurity can be localized. We solve the Schr\"{o}dinger equation and characterize the impurity by the energy spectrum of the bound states, their wave functions, and the mean position. We find the condition for the appearance of the bound states, which is a threshold for a single dimensionless parameter that involves the masses of the impurity and of the particles of the Bose gas as well as the interaction strengths. 
We note that a related phenomenon of self-trapping, i.e., the localization of a single impurity in homogeneous Bose-Einstein condensates was studied in Refs.~\cite{PhysRevB.46.301,PhysRevLett.96.210401,PhysRevA.73.063604,PhysRevA.73.043608,Bruderer_2008,ardila_strong_2020}. 
Contrary to the case of the self-trapping where a bound state is created due to the significant distortion of the density of the host system due to the coupling with the impurity, in our case it is the boundary that critically modifies the density. Other related phenomena include the localization of bosonic atoms by fermionic ones in attractive Bose-Fermi mixtures examined in Ref.~\cite{PhysRevLett.101.050402}. The formation of bounds states of two impurities immersed in one-dimensional liquid has been studied recently in Refs.~\cite{dehkharghani_coalescence_2018,PhysRevB.100.245419,huber_-medium_2019}, while the higher-dimensional cases are studied, e.g., in Ref.~\cite{naidon_two_2018,panochko_two-body_2019,camacho-guardian_bipolarons_2018}. We eventually mention the study of a ionic impurity in a condensate \cite{astrakharchik_ionic_2020}.

The paper is organized as follows. In Sec.~\ref{sec:model} we introduce the model of interacting bosons in a semi-infinite geometry. In Sec.~\ref{sec:II} we solve the equation of motion for the single-particle field operator. We find the mean-field solution and the first two quantum corrections. This enables us to evaluate the density of bosons, including its quantum contribution, in Sec.~\ref{sec:III}.  In Sec.~\ref{sec:IV} we consider the problem of a mobile quantum impurity interacting with the system. We solve the Schr\"odinger equation for the impurity and study its properties. 
Section \ref{sec:discussion} is devoted to discussions, while in Sec.~\ref{summary} we summarize our results. In Appendix \ref{appendix} we present a simplified procedure that leads to the density in the regime of large separations from the boundary.

\section{The semi-infinite Bose gas\label{sec:model}}

We are interested in the influence of the boundary at $x=0$ on the physical quantities in a one-dimensional system of interacting bosons. We thus study a long system described by the Hamiltonian
\begin{align}
H=\int_0^L dx\left(-\hat\Psi^\da \dfrac{\hbar^2\partial_x^2}{2m}\hat\Psi+\dfrac{g}{2}\hat\Psi^\da\hat\Psi^\da\hat\Psi\hat\Psi\right).
\label{H}
\end{align}
Here $m$ is the mass of bosons, while the coupling constant $g>0$ describes the repulsive contact interaction between them. The system size is $L$; however we will eventually consider the thermodynamic limit. The bosonic single particle operators $\hat\Psi$ and $\hat\Psi^\dagger$ satisfy the usual equal time commutation relations $[\hat\Psi(x,t),\hat\Psi^\da(x',t)]=\delta(x-x')$, while the other commutators vanish. The model (\ref{H}) is characterized by the dimensionless parameter \cite{lieb_exact_1963} $\gamma=mg/\hbar^2n_0$, where $n_0$ is the mean particle density. The boundary of the system imposes the nullification of the single-particle operator at its position,
\begin{align}
\hat\Psi(x=0,t)=0. \label{bc}
\end{align}
The condition (\ref{bc}) implies that the boson density also vanishes at $x=0$. 

Our first goal in this work is to calculate the density profile of the bosons 
\begin{align}\label{ndef}
n(x)=\la \hat\Psi^\da(x)\hat\Psi(x)\ra,
\end{align}
where the average is with respect to the ground state of the Hamiltonian (\ref{H}). For simplicity, we introduce the dimensionless coordinates for the position and the time, respectively, defined as
\begin{gather}
X=\frac{x}{\xi_\mu},\quad T=\frac{t}{\hbar/\mu}.
\end{gather}
Here $\xi_\mu=\hbar/\sqrt{m\mu}$ denotes the healing length, while $\mu$ is the chemical potential. Assuming the single particle operator in the form
\begin{align}
\hat\Psi(x,t)=\sqrt{\frac{\mu}{g}}\hat\psi(X,T)e^{-iT},
\end{align}
its equation of motion $i\hbar \partial_t\hat\Psi=[\hat\Psi,H]$ in the dimensionless units becomes
\begin{align}
i\partial_T\hat\psi(X,T)=\left[-\dfrac{\partial_X^2}{2}-1+\hat\psi^\da(X,T)\hat\psi(X,T)\right]\hat\psi(X,T).\label{eom}
\end{align}
We will solve \eq{eom} at $\gamma\ll 1$, corresponding to the regime of weak interaction. In this case, one can expand the field operator as \cite{sykes_drag_2009,pitaevskii_bose-einstein_2003}
\begin{align}
\hat\psi(X,T)=\psi_0(X)+\alpha \hat\psi_1(X,T)+\alpha^2\hat\psi_2(X,T)+\ldots,\label{exp}
\end{align}
such that $[\hat\psi(X,T),\hat\psi^\da(X',T)]=\alpha^2\delta(X-X')$. Here the small parameter is $\alpha=(\gamma g n_0/\mu)^{1/4}\approx \gamma^{1/4}\ll 1$. In the latter estimate we used the expression $\mu=g n_0$, which is valid at weak interaction, $\gamma\ll 1$.

The function $\psi_0(X)$ describes the time-independent wave function of the system in the absence of fluctuations. The field operators $\hat\psi_1$ and $\hat\psi_2$ account for its first and the second quantum correction. Note that the Bose-Einstein condensate does not exist in one dimension in the thermodynamic limit due to strong effect of long-wavelength fluctuations. However, in finite-size systems, the inverse system size provides an infrared cutoff. The perturbative expansion (\ref{exp}) is justified as long as \cite{sykes_drag_2009,petrov_low-dimensional_2004} $\ln(L/\xi_\mu)\ll1/\sqrt{\gamma}$ where $L$ is the length of the system. The latter inequality shows that for weakly interaction bosons [i.e., at $\gamma\ll1$] the system size can actually be huge. We point out that the density $n(x)$ [see Eq.~(\ref{ndef})] which is to be calculated, is well defined (i.e., cutoff--independent) in the thermodynamic limit \cite{Casimir}.

\section{Solution of the equation of motion\label{sec:II}}

\subsection{Wave function in the absence of fluctuations}

The form (\ref{exp}) substituted into Eq.~(\ref{eom}) leads to a hierarchy of equations controlled by the small parameter $\alpha\ll 1$. The equation of motion for $\psi_0(X)$ is obtained at order $\alpha^0$. It reads
\begin{align}
{\widehat{\mathcal{L}}}_1(X)\psi_0(X)=0,\label{eom0}
\end{align}
where we have introduced the operator
\begin{align}\label{operator}
{\widehat{\mathcal{L}}}_j(X)=-\frac{\partial^2_X}{2}+j |\psi_0(X)|^2-1.
\end{align}
The expression (\ref{eom0}) is known as the Gross-Pitaevskii equation \cite{pitaevskii_bose-einstein_2003}. This second order differential equation should be supplemented by two boundary conditions. One of them follows from Eq.~(\ref{bc}) and becomes $\psi_0(0)=0$. The other condition arises from the physical requirement that the density, which is proportional to $|\psi_0(X)|^2$, is unaffected by the boundary at long separations from it and thus becomes a constant. The real solution \footnote{Assuming that the gradient of the phase of the complex function $\psi_0(X)$ vanishes at large $X$, one concludes that the phase is a constant.} of Eq.~(\ref{eom0}) satisfying such boundary conditions is
\begin{align}
\psi_0(X)=\tanh X.\label{p0}
\end{align}
At weak interaction the chemical potential of the Bose gas is $\mu=g n_0$. This gives the density (\ref{ndef}) at the mean-field level to be
\begin{align}\label{1122}
n(x)=n_0 \tanh^2(x/\xi),\quad \xi=1/n_0\sqrt{\gamma}.
\end{align}
This expression shows that the density quickly saturates at distances beyond the healing length $\xi$ (which is $\xi_{\mu}$ in the limit $\gamma\ll 1$), provided one takes only the leading order term from the expansion (\ref{exp}). 

\subsection{First quantum correction}

We now consider the effects of quantum fluctuations and determine $\hat\psi_1$. Its equation of motion is
\begin{align}
i\partial_T\hat\psi_1(X,T)={\widehat{\mathcal{L}}}_2(X)\hat\psi_1(X,T)+\psi_0(X)^2\hat\psi_1^\da(X,T). \label{eom1}
\end{align}
We seek a solution for $\hat\psi_1(X,T)$ as a superposition of excitations of energy $\epsilon_k$ using the ansatz based on Bogoliubov transformation \cite{pitaevskii_bose-einstein_2003}
\begin{align}
\hat\psi_1(X,T)=\sum_{k>0} N_k\left[u_k(X)\hat b_ke^{-i\epsilon_k T}-v^*_k(X)\hat b_k^\da e^{i\epsilon_k T}\right]. \label{p1}
\end{align}
Since Eq.~(\ref{eom1}) is linear and homogeneous, we must account for the normalization factor  $N_k$ in Eq.~(\ref{p1}), which should be determined in such a way to satisfy the proper commutation relations between $\hat\psi_1$ fields. This is discussed further below. The bosonic operators $\hat b_k$ and $\hat b_k^\da$ satisfy the standard commutation relations $[\hat b_k,\hat b^\da_{k'}]=\delta_{k,k'}$ and $[\hat b_k,\hat b_{k'}]=0$. The boundary condition (\ref{bc}) at order $\alpha$ is given in terms of $u_k$ and $v_k$ by
\begin{align}\label{bcuv}
u_k(0)=v_k(0)=0.
\end{align}
Substitution of \eq{p1} into \eq{eom1} leads to two coupled equations for $u_k$ and $v_k$. They are known as the Bogoliubov-de Gennes equations and are given by
\begin{subequations}
\label{eq:bdguv1+2}
\begin{align}
\epsilon_k u_k(X)={}&{\widehat{\mathcal{L}}}_2(X)u_k(X)-\psi_0^2(X)v_k(X), \label{eq:bdguv1}\\
-\epsilon_kv_k(X)={}&{\widehat{\mathcal{L}}}_2(X) v_k(X)-\psi_0^2(X)u_k(X). \label{eq:bdguv2}
\end{align}
\end{subequations}
In order to simplify them, we introduce the functions $S(k,X)=u_k(X)+v_k(X)$ and $D(k,X)=u_k(X)-v_k(X)$. This enables us to decouple the equations and lead to the fourth order differential equation
\begin{subequations}
\label{eq4}
\begin{align}
\epsilon_k^2S(k,X)={\widehat{\mathcal{L}}}_3(X)\widehat{{\mathcal{L}}}_1(X) S(k,X).
\end{align}
Note that $D$ is given in terms of $S$ as
\begin{align}\label{Dk}
D(k,X)=\dfrac{1}{\epsilon_k}\widehat{\mathcal{L}}_1(X) S(k,X).
\end{align}
\end{subequations}
Four independent solutions of the forth-order equations (\ref{eq4}) are \cite{kovrizhin_exact_2001}
\begin{subequations}
\begin{gather}
S_n(X)=(-i k_n+2 \tanh X)e^{i k_n X},\label{S}\\
D_n(X)=-\dfrac{ik_n}{\epsilon_{k_n}}\frac{1}{\cosh^{2}X}e^{i k_n X}+\dfrac{k_n^2}{2\epsilon_{k_n}}S_n(X),\label{D}
\end{gather}
\end{subequations}
where $n\in \{1,2,3,4\}$, while  $\epsilon_k=\sqrt{k^2+k^4/4}$ is the energy dispersion. The four roots entering $S_n$ are $k_{1,2}=\pm k$ and $k_{3,4}=\pm i\sqrt{4+k^2}$ in terms of $k=\sqrt{2}\sqrt{\sqrt{\epsilon_k^2+1}-1}$.
The general solution of Eqs.~(\ref{eq4})  is a linear combination
\begin{subequations}
\begin{align}
S(k,X)={}&A S_1(k,X)+S_2(k,X)+B S_3(k,X),\\
D(k,X)={}&A D_1(k,X)+D_2(k,X)+B D_3(k,X).
\end{align}
\end{subequations}
Note that $S_4$ and $D_4$ do not appear since they diverge at large $X$. The two unknown coefficient $A$ and $B$ are determined using the boundary condition (\ref{bcuv}) that is terms of $S$ and $D$ becomes $S(k,0)=D(k,0)=0$. We find $A=1$ and $B=0$.
Therefore the general solution of  Eqs.~(\ref{eq:bdguv1+2}) that satisfy the boundary condition (\ref{bcuv}) are the real functions
\begin{subequations}\label{uk+vk}
\begin{align}
u_k(X)={} & k\left(1+ \dfrac{k^2+2\cosh^{-2}X}{2\epsilon_k} \right)\sin(kX) \notag\\
	&+2\left(1+ \dfrac{k^2}{2\epsilon_k}\right)\cos(k X)\tanh X,\label{uk}\\
v_k(X)={} &k\left(1- \dfrac{k^2+2\cosh^{-2}X}{2\epsilon_k} \right)\sin(kX) \notag\\
	&+2\left(1- \dfrac{k^2}{2\epsilon_k}\right)\cos(k X)\tanh X.\label{vk}
\end{align}
\end{subequations}
The normalization in Eq.~(\ref{p1}) is obtained by requiring  \cite{pitaevskii_bose-einstein_2003} 
\begin{align}
N_kN_q\int_0^{\frac{L}{\xi_\mu}} dX[u_k(X)u_q(X)-v_k(X)v_q(X)]=\delta_{k,q}.
\end{align}
This leads to $N_k=(\xi_\mu/4L\epsilon_k)^{1/2}$ at $L\gg \xi_{\mu}$. One can then verify the equal time commutation relation $[\hat\psi_1(X),\hat\psi_1(Y)]=0$. The evaluation of the other commutation relation is more involved:
\begin{align}
[\hat\psi_1(X),&\hat\psi_1^\dagger(Y)]=\sum_{k>0} N_k^2[u_k(X)u_k(Y)-v_k(X)v_k(Y)]\notag\\
={}&\frac{\xi_\mu}{L}\sum_{k>0}[\cos\boldsymbol{(}k(X-Y)\boldsymbol{)}-\cos\boldsymbol{(}k(X+Y)\boldsymbol{)}]\notag\\
={}&\delta(X-Y),\label{commrel}
\end{align}
since $\delta(X+Y)$ always equals zero for $X,Y>0$. The second equality in Eq.~(\ref{commrel}) is obtained after performing the integral over $k$ of the function $[u_k(X)u_k(Y)-v_k(X)v_k(Y)]/4\epsilon_k-2\sin(kX)\sin(kY)$, which turns out to be zero. The remaining part after integration over $k$ then gives the delta function in Eq.~(\ref{commrel}). There we use $\sum_{k>0}(\cdots)=(L/\pi\xi_\mu)\int_0^{\infty} dk(\cdots)$.

\subsection{Second quantum correction}

The first quantum contribution to the density (\ref{ndef}) is proportional to $\alpha^2$ and thus is determined by the first two corrections of the field operator in Eq.~(\ref{exp}). We thus now consider the second quantum correction to the field operator $\hat\psi$ denoted by $\hat\psi_2$. Its equation of motion is obtained from \eq{eom} at order $\alpha^2$. Since $\psi_0(X)$ is real [see Eq.~(\ref{p0})], it is sufficient to consider the real part of the expectation value  $\la\hat\psi_2\ra$ which enters into $n(x)$. For this purpose we introduce the notation $\psi_2=\text{Re}\la\hat\psi_2\ra$. The equation of motion for $\psi_2$ is
\begin{align}
\widehat{\mathcal L}_3(X) \psi_2(X)=f(X),\label{eom2}
\end{align}
where 
\begin{gather}
f(X)=-2\psi_0\la\psi_1^\da\psi_1\ra-\psi_0\la\psi_1^2\ra.\label{f}
\end{gather}
At zero temperature one has
\begin{align}
\la\psi_1^\da\psi_1\ra= \sum_{k>\lambda}  N_k^2 v_k^2 , \quad
\la\psi_1^2\ra=- \sum_{k>\lambda}  N_k^2 u_kv_k. \label{eq:pdp}
\end{align}
Note that the source term $f(X)$ in Eq.~(\ref{eom2}) is time-independent and thus $\psi_2$ is only a function of $X$. We note that evaluation of $f(X)$ requires a small-$k$ cutoff ($\lambda\sim \xi_\mu/L$) since it is divergent at $k\to 0$. 

The solution of the linear equation (\ref{eom2}) can be expressed as
	\begin{align}
		\psi_2(X)=\int_0^\infty dY\, \mathcal G(X,Y)f(Y), \label{p2}
	\end{align}
where $ \mathcal G$ is the Green's function of the operator $ \widehat{\mathcal L}_3(X)$. It satisfies
	\begin{align}
		\widehat{\mathcal{L}}_3(X) \mathcal{G}(X,Y)=\delta(X-Y).
		\label{eq:LG}
	\end{align}
Moreover, the Green's function is symmetric $\mathcal G(X,Y)=\mathcal G(Y,X)$ and satisfies
\begin{subequations}
\begin{gather}
		\mathcal{G}(Y^+,Y)=\mathcal{G}(Y^-,Y),\\
		\partial_X\mathcal{G}(X=Y^+,Y)-\partial_X\mathcal{G}(X=Y^-,Y)=-2,
\end{gather}
\end{subequations}
to account for the  $\delta(X-Y)$ on the right hand side of \eq{eq:LG}. Here $Y^\pm=\lim_{\delta\to 0^+} Y\pm \delta$. Also, the Green's function obeys $\mathcal G(0,Y)=0$ due to the boundary condition imposed by the end of the system, i.e., $\psi_2(0)=0$.
For the solution of \eq{eq:LG} we find 
\begin{subequations}\label{green+}
\begin{gather}\label{green}
\mathcal G(X,Y)=\frac{g(Y)\theta(X-Y)+g(X)\theta(Y-X)}{\cosh^2X\cosh^2Y},
\end{gather}
where
\begin{gather}
g(Y)=\frac{12Y+8\sinh(2Y)+\sinh(4Y)}{16}.
\end{gather}
\end{subequations}
Therefore, the solution of Eq.~(\ref{eom2}) for $\psi_2(X)$ is obtained from Eq.~(\ref{p2}), where one should substitute the Green's function and the source function (\ref{f}). Since the latter depends on the infrared cutoff, $\psi_2$ has also this feature. In Appendix \ref{appendix} we derive $\psi_2(X)$ in the regime $X\gg 1$.

\section{Local density\label{sec:III}}

Having solved the equation of motion for the terms $\psi_0$, $\hat\psi_1$, and $\hat\psi_2$ of the single-particle operator (\ref{exp}), we are now in position to evaluate the spatial density profile of the semi-infinite Bose gas. It is given by the expansion
\begin{align}
	n(x)=\dfrac{\mu}{g}\left[n^{(0)}(X)+\alpha^2 n^{(1)}(X) + {O}(\alpha^3)\right]{\bigg{|}_{X=\frac{x}{\xi_\mu}}}.\label{hatn}
\end{align}
Here the mean-field contribution is
\begin{gather}\label{classical}
		 n^{(0)}(X)=|\psi_0(X)|^2=\tanh^2X,
\end{gather}		 
while the quantum contribution to the density has the form
\begin{align}
		 n^{(1)}(X)=\la\hat\psi_1^\da(X,T)\hat\psi_1(X,T)\ra+2\psi_0(X)\psi_2(X), \label{nalpha}
\end{align}
or more explicitly
\begin{align}
&n^{(1)}(X)= \int_0^\infty \frac{dk}{4\pi\epsilon_k}\biggl\{v_k(X)^2 -2\tanh X \notag \\
&\times\int_0^\infty dY \mathcal G(X,Y) v_k(Y)\left[2v_k(Y)-u_k(Y)\right]\tanh Y \biggr\}. \label{n1detail}
\end{align}
In Eq.~(\ref{n1detail}) one should use Eqs.~(\ref{uk+vk}) and (\ref{green+}), while we recall $\epsilon_k=\sqrt{k^2+k^4/4}$. Unlike $\psi_2$ that requires an infrared cutoff, the density contribution (\ref{n1detail}) does not. Each of the two terms in Eq.~(\ref{nalpha}) is divergent, however their sum is finite. This should be the case since the density fluctuations are expected to be finite unlike the fluctuations of the phase of the single particle operator \footnote{In Appendix of Ref.~\cite{Casimir} is presented an alternative approach that uses the density-phase representation of the single-particle field operator, which does not rely on the expansion (\ref{exp}) and leads to the same final result.}. 

The evaluation of \eq{n1detail} is rather tedious. We first perform the integration over $Y$ using various trigonometric identities and the integration by parts. As a result, we obtain a cumbersome expression involving several hypergeometric functions. Such expression is an integrable function of $k$. To perform the integration we split the integrand into several summands. We should stress that, unlike the whole integrand, the summands can be nonintegrable due to singularities and  one should regularize them by adding and subtracting some functions. As a result of such procedure, we obtain the local density that can be expressed in the form
\begin{widetext}
\begin{align}
n(x)=\frac{\mu}{g}\left\{\tanh^2(x/\xi_\mu)+\alpha^2 \left[\frac{\tanh^2 (x/\xi_\mu)}{\pi}+\frac{(x/\xi_\mu)\tanh(x/\xi_\mu)}{\pi \cosh^2(x/\xi_\mu)}+h(x/\xi_\mu)\right]\right\}, \label{denmu}
\end{align}
where $\xi_\mu=\hbar/\sqrt{m\mu}$ and 
\begin{align}
h(z)={}&\dfrac{2}{\pi}-\dfrac{3-z\tanh z}{\pi \cosh^2  z}+\dfrac{\pi z(2-\cosh 2z)}{8\cosh^4z}\left[\dfrac{8z}{\pi^2}  \,_2F_3\left(1,1;1/2,3/2,2;4z^2 \right)-I_1(4z){\boldsymbol{ L}}_0(4z)+I_0(4z){\boldsymbol{ L}}_1(4z) \right]\notag\\
&+\tanh z(1-4z\tanh z)\left[I_0(4z)- {\boldsymbol{ L}}_0(4z)\right]-\dfrac{1}{2}[1-z \tanh z(3+5 \tanh^2 z)]\left[I_1(4z)- {\boldsymbol{ L}}_{-1}(4z)\right]. \label{den1}
\end{align}
\end{widetext}
Here $_2F_3$ is the hypergeometric function while $I_\nu(z)$ and $\boldsymbol{ L}_\nu(z)$ are the modified Bessel function of the first kind and the modified Struve function, respectively.

The density (\ref{denmu}) is obtained for a fixed value of the chemical potential. We eliminate $\mu$ by using the mean density $n_0$ that is given by $n_0=\int_0^L  dx\, n(x)/L$.
In the thermodynamic limit, the mean density of the Bose gas is $n_0=(\mu/g)(1+\alpha^2/\pi+\ldots)$. Inverting this relation, one obtains the chemical potential as a function of the mean density \cite{lieb_exact_1963}, $\mu= g n_0(1-\sqrt{\gamma}/\pi+\ldots)$, where $\gamma=mg/\hbar^2 n_0\ll 1$. Substituting $\mu$ as a function of $n_0$ in \eq{hatn} yields the density profile of the weakly-interacting semi-infinite Bose gas 
\begin{align}
n(x)=n_0\left[\tanh^2(x/\xi)+\sqrt{\gamma} h(x/\xi)\right], \label{den}
\end{align}
where $\xi=1/n_0\sqrt{\gamma}$, while $h$ is given by Eq.~(\ref{den1}).

\begin{figure}
\includegraphics[width=\columnwidth]{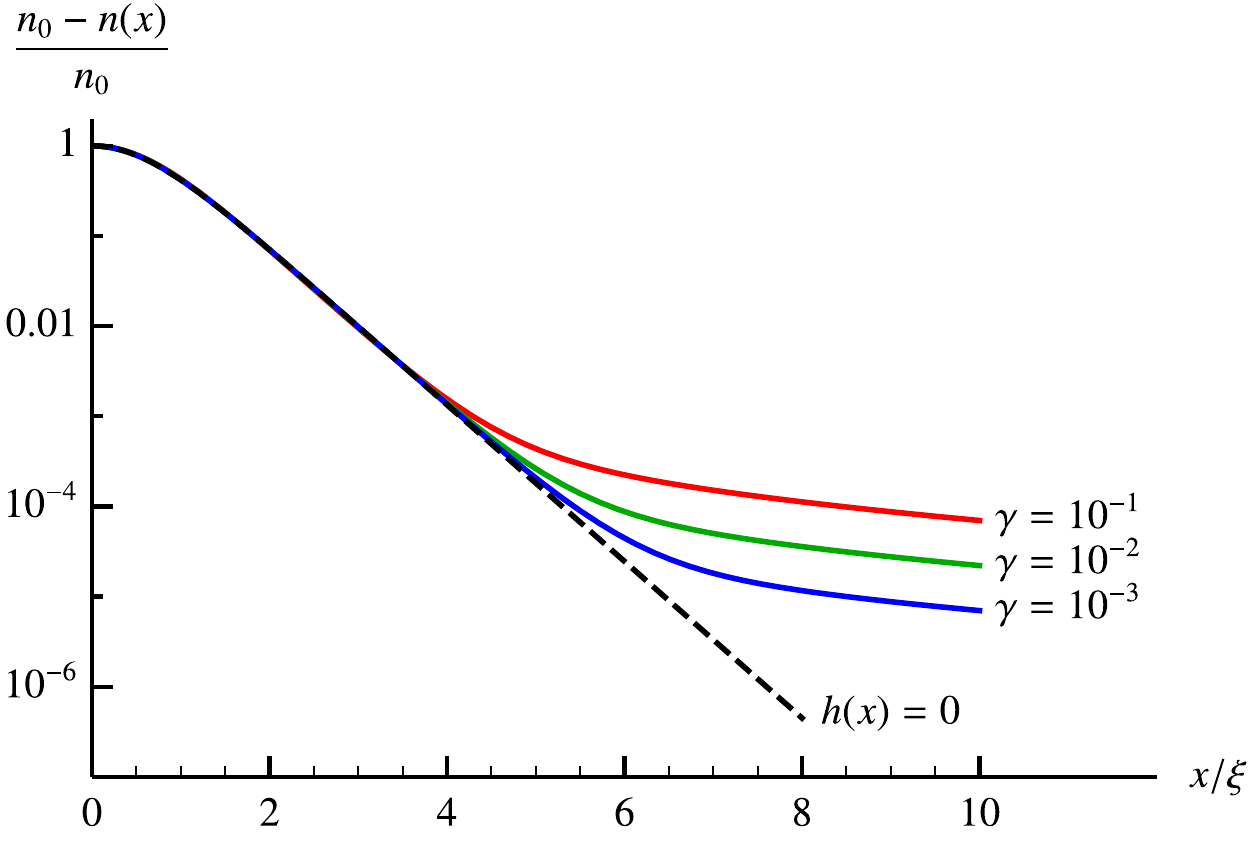}
\caption{Density deviation from the mean density $[n_0-n(x)]/n_0$ of the Bose gas for three values of the dimensionless interaction parameter $\gamma$. The dashed curve denotes the mean-field result.}
\label{fig1z}
\end{figure}

We now analyze the behavior of the local density (\ref{den}). At short separations, $x\ll \xi$, the quantum fluctuation contributions to the density can be neglected and  the leading contribution is given by
\begin{align}
n(x)=n_0\,{\tanh^2(x/\xi)}. \label{sd}
\end{align}
This classical result originates from the mean-field contribution (\ref{classical}) to the density. At longer separations, $x\gg \xi$, the classical result (\ref{sd}) quickly saturates to a constant $n_0$. However, the approach of the local density toward $n_0$ is actually much slower than one obtains from the mean-field result (\ref{sd}). At $x\gg \xi$ the quantum contribution   (\ref{nalpha}) is the dominant term in the density deviation $n(x)-n_0$ since it decays algebraically with the distance. This follows from the asymptotic expansion
\begin{align}
h(z)=-\dfrac{1}{16 \pi z^2}\left[1+\dfrac{1}{z}+\dfrac{15}{16z^2}+O(z^{-3}) \right].\label{abc}
\end{align}
Therefore, at $x\gg \xi$ the leading correction to the local density is given by
\begin{align}
n(x)=n_0\left(1-\frac{\sqrt{\gamma}}{16\pi}\frac{\xi^2}{x^2}\right).\label{ld}
\end{align}
The crossover distance $x_c$ between the two regimes can be estimated by equating the two expressions (\ref{sd}) and (\ref{ld}). As a result we get
\begin{align}\label{eq:xc}
x_c \approx \xi \ln\left(\frac{8\sqrt{\pi}}{\gamma^{1/4}}\right).
\end{align}
Note that the crossover distance depends only logarithmically on $\gamma$ and practically is on the order of few $\xi$. This is illustrated in Fig.~\ref{fig1z} where  the density profile of the Bose gas is shown as a function of the distance from the end of the system at several values of $\gamma$. The density in the crossover regime between the classical and the quantum one is given by Eq.~(\ref{den}). A simplified derivation of $n^{(1)}(X)$ at $X\gg 1$ is given in Appendix \ref{appendix}. We finally notice that the ground-state energy calculation of Ref.~\cite{PhysRevB.100.235431} leads to the limiting forms (\ref{sd}) and (\ref{ld}).

\section{Interaction-induced localization of a single impurity\label{sec:IV}}

In this section we study the problem of a quantum impurity in the semi-infinite Bose gas. We consider the impurity that is locally coupled to the boson density (\ref{den}). In the case of repulsion, the density profile of the Bose gas forms an attractive potential for the impurity, which can lead to the bound states in the spectrum. In that case the particle becomes localized by the surrounding medium.

In the case of a weakly coupled impurity, one can find its wave function in the approximation of unperturbed Bose gas density (as further discussed in Sec.~\ref{sec:discussion}). The impurity wave function  is a function of $x$ ant $t$ and is governed by the Schr\"{o}dinger equation. After separation of variables it reduces to the eigenvalue problem on the positive semi-axis $x>0$
\begin{subequations}\label{SCHR+zbc}
	\begin{gather}\label{SCHR}
\left[-\frac{\hbar^2}{2M}\frac{d^2}{dx^2}+G n(x) -E\right]\psi_{\rm imp}(x)=0,\\
\psi_{\rm imp}(0)=0.\label{zbc}
\end{gather} 
\end{subequations}
Here $M$ is the impurity mass, $G>0$ denotes the coupling constant, while the density of the Bose gas $n(x)$ is given by Eq.~(\ref{den}). An alternative formulation of the eigenvalue problem (\ref{SCHR+zbc}) is to consider unrestricted $x$ with the symmetric potential $Gn(|x|)$, and to account only for odd eigenfunctions. They have a node at $x=0$ and thus automatically satisfy the boundary condition (\ref{zbc}). 

In order to simplify the notations, we rewrite Eq.~(\ref{SCHR+zbc}) in the form
\begin{subequations}
	\label{eomff}
\begin{gather}\label{eomffa}
\left[-\frac{d^2}{dz^2}-\frac{\lambda(\lambda-1)}{\cosh^2 z}+\sqrt{\gamma}\lambda(\lambda-1)h(z) +\varkappa^2\right]f(z)=0,\\
f(0)=0,
\end{gather}
\end{subequations}
where we have introduced
\begin{gather}
f(z)=\psi_{\rm imp}(z \xi), \label{n1}\\
\lambda=\dfrac{1}{2}+\dfrac{1}{2}\sqrt{1+\dfrac{8G M}{g m}},\label{n2}\\
\varkappa^2=\frac{2GM}{gm}\left(1-\frac{E}{G n_0}\right).\label{n3}
\end{gather}
In the following we will find the bound state spectrum for the eigenvalue problem (\ref{eomff}).

\subsection{Bound states of P\"{o}schl-Teller potential}

In the limit where the quantum correction to the density $h(z)$ is neglected, the potential of Eq.~(\ref{eomffa}) describes a hole of modified P\"{o}schl-Teller type which admits an exact solution \cite{pschl_bemerkungen_1933,flugge_practical_2012}.
We now derive the bound states for this special eigenproblem defined by
\begin{subequations}
	\label{f0}
	\begin{gather}\label{f0a}
	\left[-\frac{d^2}{dz^2}-\frac{\lambda(\lambda-1)}{\cosh^2 z}+\varkappa^2\right]f(z)=0,\\
	f(0)=0.
	\end{gather}
\end{subequations}
From the definition (\ref{n2}) follows $\lambda>1$ and thus the potential $-\lambda(\lambda-1)/\cosh^2 z$ in Eq.~(\ref{f0a}) is negative. We also notice that for unrestricted $z$, the eigenstates in such symmetric potential  can be even or odd functions of $z$. For the problem (\ref{f0}) only odd solutions have a node at $z=0$. They are of the form  \cite{flugge_practical_2012}
\begin{align}
f(z)=&\cosh^\lambda z\sinh z \notag\\ &\times\, _2F_1\left(\frac{\lambda-\varkappa+1}{2},\frac{\lambda+\varkappa+1}{2};\frac{3}{2};-\sinh^2z\right),\label{sol:f0}
\end{align}
where $_2F_1$ is the Gauss's hypergeometric function.

In order to discuss the appearance of bound states, we consider the asymptotic expansion of \eq{sol:f0} in the limit $z\to\infty$, which is given by
\begin{align}\label{fo}
f(z)\simeq \frac{2^{-\varkappa}\Gamma(\varkappa)\Gamma(3/2)e^{\varkappa z}}{\Gamma\left(\frac{\lambda+\varkappa+1}{2}\right) \Gamma\left(\frac{2-\lambda+\varkappa}{2}\right)}+\frac{2^{\varkappa}\Gamma(-\varkappa)\Gamma(3/2)e^{-\varkappa z}}{\Gamma\left(\frac{\lambda-\varkappa+1}{2}\right) \Gamma\left(\frac{2-\lambda-\varkappa}{2}\right)}.
\end{align}
We notice that there are two exponential terms $\propto e^{\pm \varkappa z}$ at $z\to \infty$. Taking into account the definition of $\varkappa$ given by Eq.~({\ref{n3}), we conclude that a bound state can be realized only if $\varkappa^2>0$, i.e., if $G n_0>E$. In the opposite case, $Gn_0<E$, the solution at long separations from the boundary is an oscillating trigonometric function since $\varkappa$ is purely imaginary. In the following we focus on the localized (bound) states and derive their energy spectrum. We assume without loss of generality that $\varkappa>0$. Thus, the coefficient in front of the term $e^{\varkappa z}$ must vanish in order to obtain a nondivergent (normalizable) wave function at large $z$. This occurs when the arguments of the Gamma functions in the denominators are 0 or negative integers. Since $\lambda>1$ and by assumption $\varkappa>0$, only the argument of $\Gamma\left(\frac{2-\lambda+\varkappa}{2}\right)$ matters. This yields
\begin{align}
\varkappa=\lambda-2-2\eta, \label{bs}
\end{align}
where $\eta=0,1,2,\ldots $. From the condition $\varkappa>0$ we obtain that integer $\eta$ satisfies
\begin{align}
0\le \eta< {\frac{\lambda}{2}-1}.\label{etacond}
\end{align}
Equation (\ref{etacond}) determines the condition $\lambda>2$ necessary for the appearance of the first bound state \footnote{Unlike the potential $-\lambda(\lambda-1)/\cosh^2 z$ that has bound states whenever is negative (i.e., for $\lambda>1$), in our setup such potential is present only for $z>0$, while at $z<0$ the potential is infinite. This renders the nullification of the wave function at the origin. As a consequence, the lowest bound state occurs for $\lambda>2$, i.e., when the potential is sufficiently deep.}. The number $\eta$ denotes the number of nodes of the odd bound-state eigenfunction in the region of interest $z>0$. Thus, the localized solutions of the eigenvalue problem \eq{f0} are given by Eq.~(\ref{sol:f0}) where $\varkappa$ satisfies \eq{bs}, while $\eta$ is a nonnegative integer satisfying the condition (\ref{etacond}). The latter enables us to reexpress the eigenfunction (\ref{sol:f0}) in the form
\begin{align}
f_{\eta}(z)={}&\eta!\, \Gamma(\lambda-\eta-1)\frac{\sinh z}{\cosh^{\lambda-1}z}\notag\\
	&\times\sum_{j=0}^\eta \dfrac{(-4)^j\sinh^{2j}z}{(2j+1)!\, (\eta-j)!\,\Gamma(\lambda-\eta-1-j)}\label{fz}
\end{align}
using $_2F_1(c-a,c-b;c;z)=(1-z)^{a+b-c}\, _2F_1(a,b;c;z)$.

The preceding  expressions enable us to find the spectrum of bound states of the impurity defined by Eq.~(\ref{SCHR+zbc}) in the case when the quantum fluctuations of the density are neglected. The eigenfunctions are $\psi_{\rm imp,\eta}(x)=f_\eta(x/\xi)$ [see Eq.~(\ref{fz})], where one should substitute Eq.~(\ref{n2}) for $\lambda$. The corresponding energies [cf.~Eq.~(\ref{bs})] are
\begin{align}\label{energy}
	E_\eta=g n_0 \dfrac{m }{M}\left[\left(\dfrac{3}{4}+\eta\right)\sqrt{1+\dfrac{8 G M}{g m}}-\dfrac{5}{4}-3\eta-2\eta^2\right],
\end{align}
provided the system has bound states. The condition for that is sufficiently strong coupling between the impurity and the Bose gas, which should satisfy
\begin{align}\label{condition}
\frac{GM}{gm}>(\eta+1)(2\eta+1).
\end{align}
The latter follows from Eq.~(\ref{etacond}). The set of allowed values for $\eta$ (non-negative integers) in the spectrum (\ref{energy}) are determined by the inequality (\ref{condition}). In particular, the first bound state (corresponding to $\eta=0$) occurs for ${GM}/{gm}>1$. The second one (corresponding to $\eta=1$) exists for ${GM}/{gm}>6$, etc. We introduce an important quantity
\begin{align}\label{binding}
E^{b}_{\eta}=G n_0-E_{\eta},
\end{align}
which denotes the binding energy, $E^{b}_{\eta}>0$.

In Fig.~\ref{fig2} we show the bound state energies for the first five levels. We notice that the specific bound state energy expressed in units of $G n_0$, which is the value of the potential at infinity, is a function of the single parameter $GM/gm$. In Fig.~\ref{fig3} we show the normalized wave functions for the bound state levels for the specific value $GM/gm=30$. We notice that the wave functions for energies near the top of the potential, $G n_0$, are weakly localized as their spatial extension increases. The level (quantum) number $\eta$ corresponds to the number of nodes of the wave function. For a given wave function, its spatial extension decreases with increasing $GM/gm$, since in that case the potential becomes deeper. This is reflected in the mean distance of the particle from the origin,
\begin{align}
\langle x\rangle_\eta=\xi  \frac{\int_0^\infty dx x |f_{\eta}(x)|^2}{\int_0^\infty dx |f_{\eta}(x)|^2}.
\end{align}
We plot this dependence in Fig.~\ref{fig4z}. Close to the threshold for the appearance of the bound state, the mean distance diverges according to the law
\begin{align}\label{trans}
\frac{\langle x\rangle_\eta }{\xi}=\frac{1}{2\varkappa}=\frac{\eta+\frac{3}{4}}{\frac{GM}{gm}-(\eta+1)(2\eta+1)}+O(1).
\end{align}
The denominator of Eq.~(\ref{trans}) denotes the distance from the threshold for the appearance of the bound state $f_{\eta}$ measured from the ``localized'' side where the particle is in the bound state. Equation (\ref{trans}) can be also expressed as function of the binding energy ${\langle x\rangle_\eta }/{\xi}=\sqrt{m\mu/8M E^b_{\eta}}$  using Eqs.~(\ref{n3}) and (\ref{binding}).
Indeed, ${\langle x\rangle_\eta }$ determines the localization length. This easily follows from the asymptotic expansion of the wave function $f_{\eta}(z)\sim \exp{(-\varkappa z)}$ [cf.~Eq.~(\ref{fo})], where the localization length is $1/\varkappa$ close to the ``transition''.  The localization length diverges with the first power of the distance from the ``transition'', which physically denotes the disappearance of the bound state as $GM/gm$ is decreased (and the binding energy approaches zero).
More generally, the universal form of the wave function enables us to find any moment as ${\langle x^j\rangle_\eta }/{\xi^j}={j!}/{(2\varkappa)^j}$ close to the threshold.

In the preceding part we considered a localized impurity, i.e., its bound states which exist when the eigenenergies are smaller than $G n_0$ (corresponding to the value of the potential at infinity). We should stress that the impurity has a continuum of scattering states with energies higher than $G n_0$. They are characterized by the oscillating wave functions [cf.~Eq.~(\ref{fo})].

\begin{figure}
		\centering
		\includegraphics[width=\columnwidth]{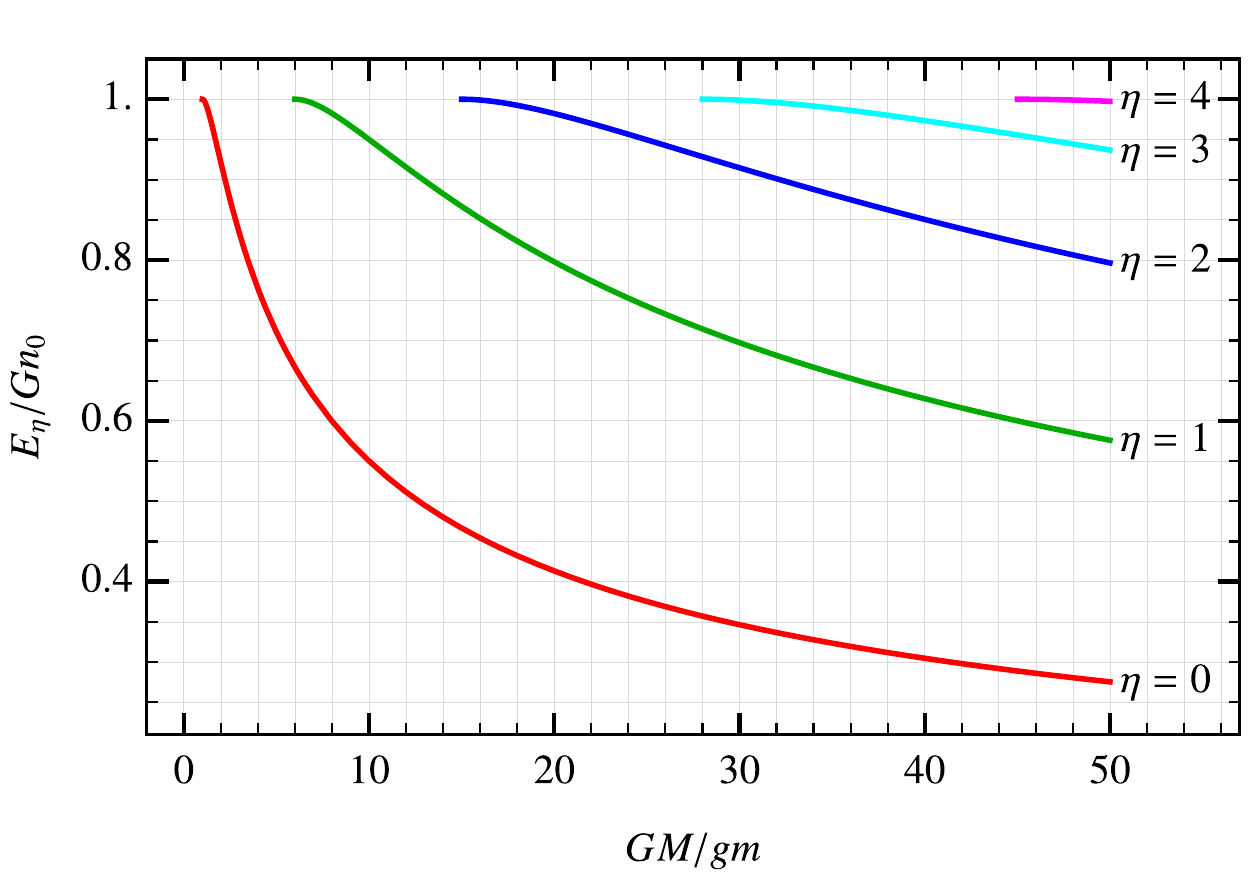}
		\caption{The first five bound state energies $E_\eta/Gn_0$ as a function of the dimensionless parameter $GM/gm$. They occur, respectively, at ratios $GM/gm=1,6,15,28,45$. Notice that the energies start from $G n_0$, which is the value of the potential at infinity.}
\label{fig2}
\end{figure} 
	
\begin{figure}
\centering
\includegraphics[width=\columnwidth]{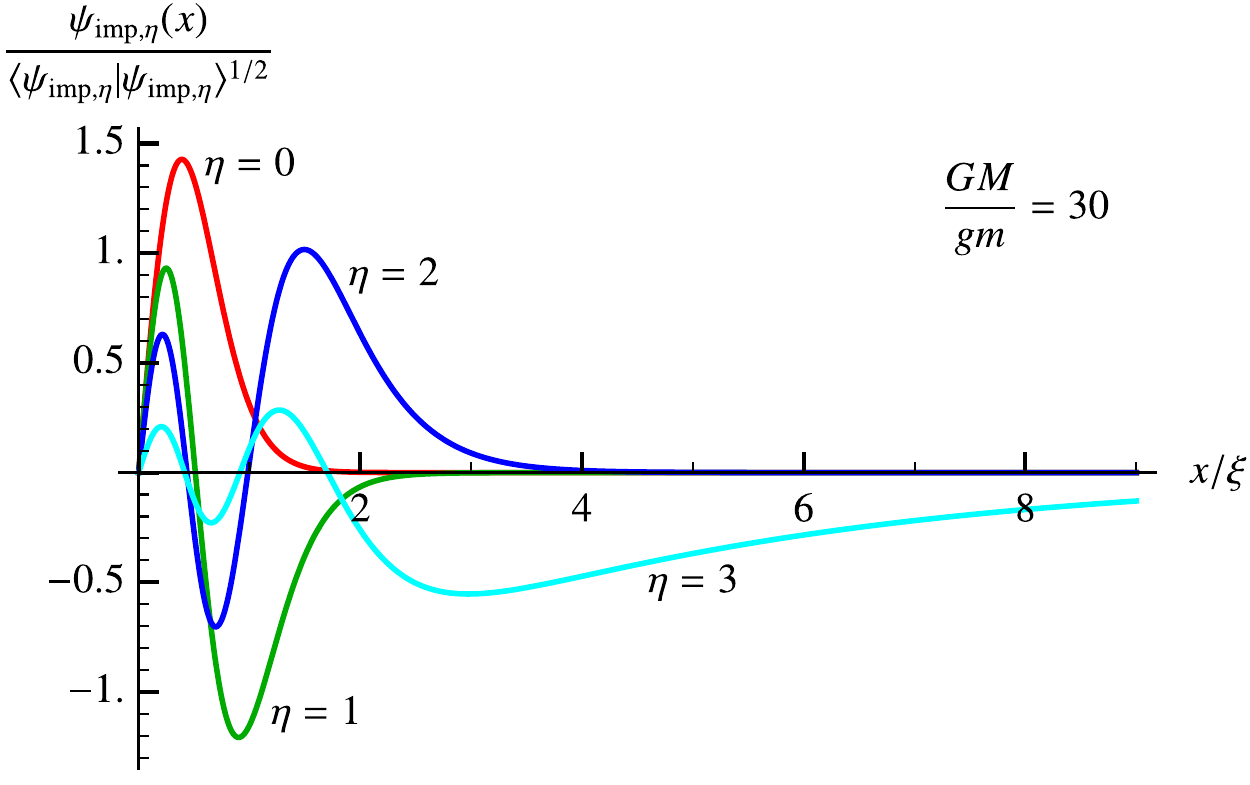}
\caption{Normalized wave functions for the first four states at $GM/gm=30$. Notice that the wave function corresponding to $\eta=3$ is spatially less localized since the value of $GM/gm$ is close to the threshold for its appearance.}
\label{fig3}
\end{figure} 	

\begin{figure}
	\centering
	\includegraphics[width=\columnwidth]{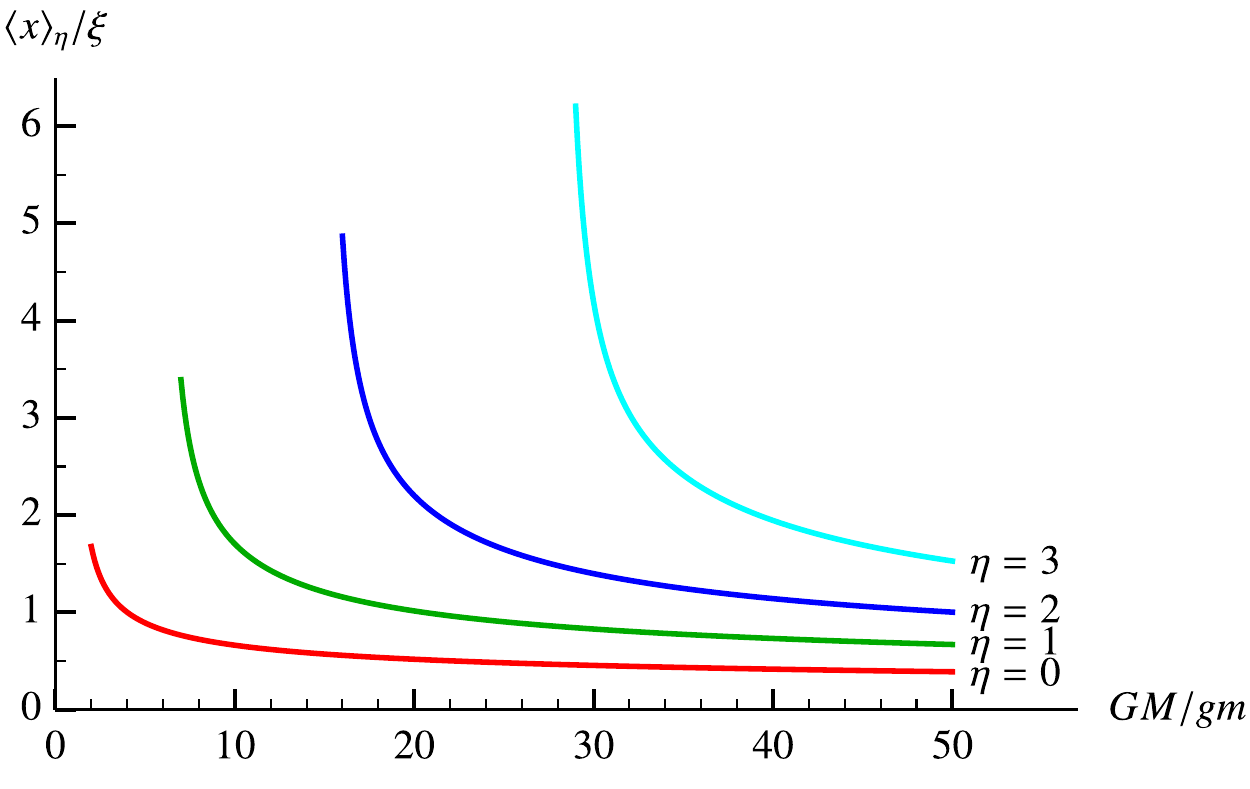}
	\caption{Mean distance $\langle x\rangle_\eta/\xi$ of the impurity from the origin for the first four bound states as a function of $GM/gm$. At the threshold value for a given $\eta$, the mean distance diverges according to Eq.~(\ref{trans}), signaling the disappearance of the bound state.}
	\label{fig4z}
\end{figure} 
	
	\subsection{Quantum correction to the potential}

In the previous subsection we have solved the eigenproblem for the impurity using the mean-field contribution to the potential. We now account for the quantum correction $h(z)$, see Eq.~(\ref{eomff}). We were not able to solve analytically the whole eigenproblem, since the form of $h(z)$ is very complicated. However we can take advantage of the small parameter $\sqrt{\gamma}$ that controls it and find the corrections of the bound state energy (\ref{energy}) using perturbation theory. The energy correction is given by	\begin{align}\label{qcorrection}
\delta E_\eta=\sqrt{\gamma}\,Gn_0\frac{\int_0^\infty dx |f_{\eta}(x)|^2h(x)}{\int_0^\infty dx |f_{\eta}(x)|^2},
\end{align}
where $h$ is defined in \eq{den1}, while $f_{\eta}$ by \eq{fz}. The quantum correction $\sqrt{\gamma}\, G n_0h(x/\xi)$ to the effective potential [see, e.g.,  Eq.~(\ref{SCHR})] becomes practically important at distances $x\gg \xi$ (see Fig.~\ref{fig1z}) with respect to the classical one $-G n_0/\cosh^2(x/\xi)$. Contrary to the  function $h(x/\xi)$ that has a long-range tail $\propto -1/x^2$, the wave function is localized and exponentially small at $x\gg \xi$. As a result the numerator in \eq{qcorrection} is small thus the quantum correction to the bound state energy is negligible. Notice that very near the threshold for the bound state appearance (see, e.g., the curve for $\eta=3$ in Fig.~\ref{fig3}) the wave function may have some overlap with the tail of $h(x/\xi)$, and the correction $\delta E_\eta$ can be somewhat important with respect to $E_\eta-G n_0$. Nevertheless, $\delta E_\eta$ is negative since the quantum correction to the potential broadens the well. The quantum corrections to the different energy levels of the bound states are numerically evaluated using Eq.~(\ref{qcorrection}) and shown in Fig.~\ref{fig5} \footnote{Since $1/x^2$ potential is rather special in quantum mechanics~\cite{essin_quantum_2006}, we performed numerical diagonalization of the full problem given by Eq.~(\ref{eomff}). We have verified that the correction term (\ref{qcorrection}) to the bound state energies are actually quite accurate approximation to the exact value, as one would naively expect without knowing about the peculiarities of $1/x^2$ potential.}.

\begin{figure}
\centering
\includegraphics[width=\columnwidth]{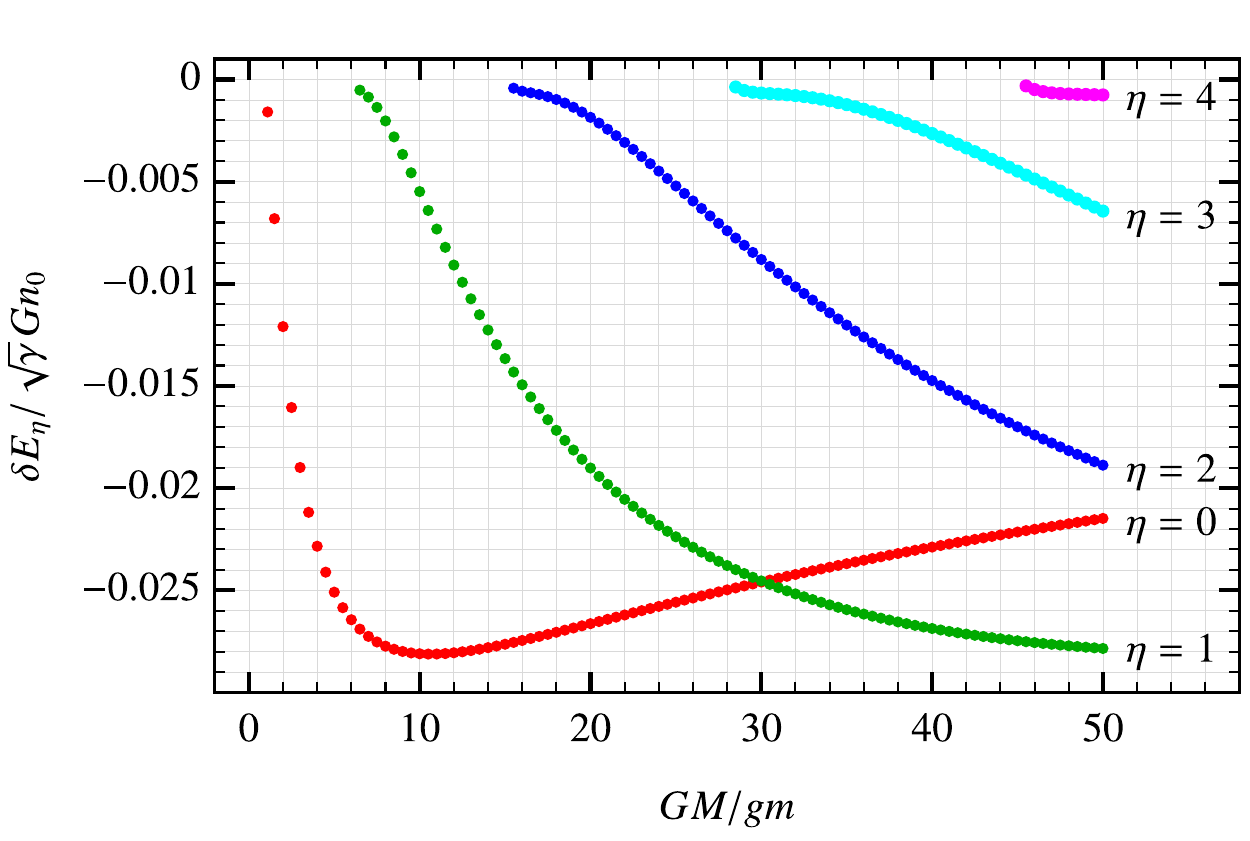}
\caption{Quantum correction to the bound state energies $\delta E_\eta/\sqrt{\gamma}\,Gn_0$ for different $\eta$ as a function of $GM/gm$ for the first five levels.}
\label{fig5}
\end{figure}

\section{Discussion\label{sec:discussion}}

We have calculated the bound states of the impurity in a semi-infinite Bose gas. We assumed that the density of the Bose gas it not affected by the presence of the impurity [cf.~Eq.~(\ref{SCHR+zbc})]. This approach is valid in cases of weak coupling $G$ between the impurity and the particles of the Bose gas. A more general model  would be the Hamiltonian (\ref{H}) supplemented by the impurity part
\begin{align}
\int_0^L dx \left(-\hat \Psi_{\rm imp} \frac{\hbar^2 \partial_x^2}{2M}\hat \Psi_{\rm imp}+ G \hat\Psi_{\rm imp}^\dagger \hat\Psi_{\rm imp} \hat{\Psi}^\dagger \hat\Psi\right).
\end{align}
From the solution of the whole problem one could obtain the spectrum of the bound states at any $G$. However this is in practice difficult to do analytically. The parameter region where our approach applies can be obtained by calculating the correction to the Bose gas density (\ref{den}) due to the coupling with the impurity. For a heavy impurity such correction is small  at $G\ll g/\sqrt{\gamma}$ \cite{PhysRevB.100.235431}. In this regime is therefore justified to study the simplified problem~(\ref{f0}). Notice that for weakly interacting bosons we have $\gamma\ll 1$ and thus our work covers a big range of values for $G$.

The quantum contribution to the boson density that universally behaves as $1/x^2$ at $x\gg \xi$ [see, e.g., Eq.~(\ref{ld})] leads to a small energy correction to the bound state levels of the localized particle. However, the knowledge of such spatial dependence of the density of particles far from the boundary is fundamentally important for quantities where the density gradient matters. An example is the Casimir-like interaction \cite{recati_casimir_2005,schecter_phonon-mediated_2014,Casimir,PRBCasimir,pavlov_phonon-mediated_2018} between the impurity and the boundary of the system mediated by density fluctuations of the medium. In the regime $x\gg \xi$ its form  was found in Ref.~\cite{PhysRevB.100.235431}, while our result (\ref{den}) determines the interaction law at all distances. It takes the form 
\begin{align}\label{ant}
U(x)={}&G [n(x)-n_0]\notag\\
={}&-G n_0\left[\frac{1}{\cosh^2(x/\xi)}-\sqrt{\gamma} h(x/\xi)\right].
\end{align}
It is expected that  under the influence of the  long-range potential (\ref{ant}), a heavy impurity immersed in the system far from the boundary slowly drifts towards it and eventually get localized. During this process, the impurity energy that equals $G n_0$ at long distances is dissipated by exciting the Bose gas.

The boundary in our system can exist naturally as the system's end. Alternatively, it can be created by a heavy impurity strongly coupled to the Bose gas that creates impenetrable potential and leads to the complete depletion of the boson density at its position. Thus the effective interaction between the boundary and the impurity can also be  interpreted as the Casimir-like interaction between two very different impurities. We notice two related very recent experimenal works \cite{desalvo_observation_2019,edri_observation_2020}, which  have demonstrated the existence of the induced interaction between quantum particles mediated by the surrounding quantum gas.

The Hamiltonian (\ref{H}) can be studied within the dual model of attractive fermions \cite{cheon_fermion-boson_1999,yukalov_fermi-bose_2005}. The density of bosons in the original model is equal to the density of fermions in the dual one. Such mapping is particularly useful in the regime of strong interaction between bosons, corresponding to $\gamma\gg 1$ and the Luttinger liquid parameter \mbox{$K\simeq 1+4/\gamma$}. In this case the fermions are weakly-attractive, which enables one to study their density, obtain the characteristic form with Friedel oscillations and go beyond the Tomonaga-Luttinger liquid description.
It is interesting to study the density and the fate of Fridel oscillations as the value of $K$ increases and find a connection with our result (\ref{den}). A numerical study of finite number of bosons \cite{hao_ground-state_2006} also suggests the above picture. Note that the model (\ref{H}) is integrable in the box geometry \cite{gaudin_boundary_1971} and thus it is in principle possible to use the exact Bethe ansatz solution to find the exact density profile in the thermodynamic limit and answer the above question. We are not aware of such study in the literature.

Recent experimental realizations of cold bosons trapped in a box potential \cite{gaunt_bose-einstein_2013,chomaz_emergence_2015,rauer_recurrences_2018,garratt_single-particle_2019} provide a possibility to observe our results. 
In order to see the characteristic long-range tail of the density [cf.~Eq.~(\ref{ld})], the system length has to be much longer than the crossover distance (\ref{eq:xc}). In addition, the constraint on the temperature comes from the fact that the thermal length $\ell_T=\hbar v/2\pi T$ has to be longer than the the crossover distance (\ref{eq:xc}). Here $v=\sqrt{g n_0/m}$ denotes the sound velocity of weakly interacting bosons. Thus we obtain $T/\mu\ll \log{(\gamma^{-1})}$.
Concerning the observation of the bound states, the condition (\ref{condition}) could be realized for a given mass ratio $M/m$ by tuning the interaction strengths using the Feshbach resonance. Temperature should be sufficiently small to avoid smearing of the energy levels. 

\section{Summary\label{summary}}

In this paper we first calculated the density profile of the weakly-interacting semi-infinite one-dimensional Bose gas. We solved the equation of motion for the field operator. At the mean-field level, it leads to the density profile (\ref{1122}). Taking into account the effect of quantum fluctuations around the mean-field solution we calculated the quantum contribution to the density, see Eqs.~(\ref{den1}) and (\ref{den}). It shows the universal $1/x^2$ behavior at long distances.

We then studied the spectrum of the bound states of a quantum impurity which experiences the potential created by the surrounding Bose gas. We were able to exactly solve this problem accounting for the mean-field boson density. The discrete spectrum is given by Eq.~(\ref{energy}), while the corresponding wave function are written in Eq.~(\ref{fz}). Bound states exist for sufficiently strong repulsion between the impurity and particles of the Bose gas, see Eq.~(\ref{condition}). We also found how the mean particle distance from the boundary behaves, and in particular its divergence (\ref{trans}) as one approaches the threshold for the appearance of the bound state levels. We showed that the quantum contribution to the density gives rise to the negligible corrections of the bound state levels. However, the quantum contribution to the density is important since it leads to the long-range Casimir-like interaction between the impurity and the boundary.

\section*{Acknowledgments}
This study has been partially supported through the EUR grant NanoX ANR-17-EURE-0009 in the framework of the ``Programme des Investissements d’Avenir''.

\appendix
\section{The density contribution at long distances\label{appendix}}

Here we present derivation of the quantum contribution to the density at long separation from the boundary. We begin with the expression for $\psi_2(X)$ that simplifies at $X\gg 1$. In order to obtain it we first notice that Eq.~(\ref{p2}) leads to the first-order differential equation
\begin{align}\label{de1}
\frac{d\psi_2(X)}{d X}={}&-2\tanh X \psi_2(X)\notag\\
&+2\cosh^2 X \int_{X}^{\infty} dY \frac{f(Y)}{\cosh^2 Y}.
\end{align}
In the regime $X\gg 1$ one can set $\tanh X$ to 1 and $\cosh^{-2} X$ to 0 in Eqs.~(\ref{uk+vk}), such that the source term becomes
\begin{align}
f(Y)={}&-\sum_{k>\lambda} N_k^2\left(1-\frac{2k^2}{\epsilon_k}+\frac{3k^4}{4\epsilon_k^2} \right) \notag\\
&\times\left[k\sin(kY)+2\cos(kY)\right]^2.
\end{align}
One can then perform the integration in Eq.~(\ref{de1}) for which it suffices to approximate $\cosh^2 Y$ by $e^{2Y}/4$ and solve an elementary integral. Solving the differential equation we then find
\begin{align}\label{psi2large}
\psi_2(X)={}&-\frac{1}{4}\sum_{k>\lambda} N_k^2\left(1-\frac{2k^2}{\epsilon_k}+\frac{3k^4}{4\epsilon_k^2} \right)\biggl[4+k^2\notag\\
&+\frac{4-k^2}{1+k^2}\cos(2kX)+\frac{4k}{1+k^2}\sin(2 k X)\biggr]
\end{align}
at $X\gg 1$ (corresponding to $x\gg \xi$). 

The quantum contribution to the density (\ref{nalpha}) can now be found more easily using Eq.~(\ref{psi2large}). It leads to
\begin{align}
n^{(1)}(X)={}&\int_0^\infty \frac{dk}{2\pi}\biggl\{1-\frac{k}{\sqrt{4+k^2}}+\biggl[\frac{5}{3(4+k^2)} \notag\\
&-\frac{2}{3(1+k^2)}-\frac{k}{(4+k^2)^{3/2}}\biggr] \notag\\
&\times\left[\left(1-\frac{k^2}{4}\right)\cos(2kX)+k \sin(2kX)\right]\biggr\}.
\end{align}
The fist two terms in braces give a constant after the integration. The next two terms contain poles in the complex plane and (after proper regularization) lead to exponentially small result. Finally the term with the  denominator with the power exponent $3/2$ is responsible for the power-law decay of $n^{(1)}(X)$. The actual evaluation yields 
\begin{align}
n^{(1)}(X)={}&\frac{3}{\pi}+\frac{3}{2}e^{-2X}-\frac{10}{3}e^{-4X}\notag\\
&+(1-4X)\left[I_0(4X)- {\boldsymbol{ L}}_0(4X)\right]\notag\\
&-\frac{1}{2}(1-8X)\left[I_1(4X)- {\boldsymbol{ L}}_{-1}(4X)\right].\label{eqeq}
\end{align} 
Equation (\ref{eqeq}) can be recognized in the density correction given by Eqs.~(\ref{denmu}) and (\ref{den1}). Apart from the exponential terms that are negligible at $X\gg 1$, the last two terms in Eq.~(\ref{eqeq}) correspond to the second line of Eq.~(\ref{den1}) taken at large argument. They lead to the power-law decay of the density correction at long separations [cf.~Eq.~(\ref{abc})].

\end{document}